# Machine learning exploration of topological polarization pattern in hexagonal boron nitride moiré superlattice

Jun-Ding Zheng[1, †], Cheng-Shi Yao[1, †], Song-Chuan Zhou[1], Yu-Ke Zhang[1], Zhi-Qiang Bao[1], Wen-Yi Tong[1], Jun-Hao Chu[1,3], Chun-Gang Duan[1,2,3,*]

[1] Key Laboratory of Polar Materials and Devices, Ministry of Education, East China Normal University, Shanghai 200241, China.

[2] Collaborative Innovation Center of Extreme Optics, Shanxi University, Taiyuan, Shanxi 030006, China.

[3] Shanghai Center of Brain-inspired Intelligent Materials and Devices, East China Normal University, Shanghai 200241, China.

[†] Equally contributed authors.

[*] Author to whom any correspondence should be addressed.

**E-mail:** cgduan@clpm.ecnu.edu.cn

## Abstract

Twisted moiré supercells, which can be approximated as a combination of sliding bilayers and constitute various topologically nontrivial polarization patterns, attract extensive attention recently. However, because of the excessive size of the moiré supercell, most studies are based on effective models and lack the results of first-principles calculation. In this work, we use machine learning to determine the topological structure of the polarization pattern in twisted and strained bilayer of hexagonal boron nitride ($h$-BN). We further confirm that the topological pattern can be effectively modulated by the vertical electric field and lattice mismatch. Finally, local polarization also exists in the antiparallel stacked $h$-BN twisted and strained bilayers. Our work provides a detailed study of the polarization pattern in the moiré superlattice, which we believe can facilitate more research in moiré ferroelectricity, topological physics, and related fields.

Twisted bilayer systems can generate long-wavelength moiré patterns, and the resulting band flattening can suppress band dispersion, thereby leading to significant many-body interactions that give rise to novel physical phenomena[1-8]. Numerous primitive cells with varying interlayer sliding can form huge moiré patterns, and the corresponding physical quantities, including excitons, magnetization, polarization, local bandgap, and interlayer potential also show a long-wavelength distribution [9-20]. Recent studies have shown that the spatial modulation of interlayer sliding can carry sliding ferroelectricity[16, 21-24]. The out-of-plane polarization supplies an effective Zeeman field, which allows layer pseudospin to be the quantum degree of freedom[19, 25, 26]. On the other hand, the in-plane and out-of-plane polarization will lead to rich topologically nontrivial patterns that could result in novel physics such as negative capacitance, high-density information processing, and various topological phenomena[27-33]. However, most of the studies on polarization patterns are based on effective models or direct mapping of local structures[34, 35], and results of first-principles calculations are lacking due to the excessive size of the moiré supercell. In recent years, machine learning schemes have provided a deep neural network, called deep Wannier (DW)[36-39], to model the centers of electronic charge via maximally localized Wannier functions (MLWFs) and Wannier centers (WCs), which allows to simulate the polarization of large moiré superlattice. In this work, we focus on the topological structure generated by sliding ferroelectricity in moiré supercell, and select the flat *h*-BN moiré supercell as the object to predict the local polarization distribution by DW and demonstrate the presence of topological nontrivial polarization patterns. Moreover, we find that the vertical electric field and lattice mismatch can effectively modulate the topological pattern (winding number). Finally, the local polarization exists even in the antiparallel stacked *h*-BN twisted bilayer, as the local structure cannot be completely equivalent to the primitive cell with interlayer sliding. We anticipate that our work can contribute to understanding and regulating moiré ferroelectricity and exploring related topological physics.

We first analyze the sliding ferroelectricity of the *h*-BN bilayer. The space group and point group of *h*-BN monolayer are *P6m2* and $D_{3h}$. Following the strategy of Ref [40], the stacking operation can be expressed as $\hat{\tau}_z \hat{O}$, where $\hat{\tau}_z$ represent the out-of-plane translation operation and $\hat{O} = \{O|\tau\}$ is transformation operator, which can be divided into two part: rotational part $O$ and in-plane translation part $\tau$. The rotational part of *h*-BN bilayer has invariant operation *E* and inversion operation *I*, corresponding to AA and AA' stacking respectively. Since the translation operation cannot break the inversion operation, the sliding ferroelectricity only occurs in $\{E|\tau\}$ operation, and the polarization as a function of $\tau$ forms a configuration space. Figure. 1(a) and (d) show the crystal structure, point group, and the direction of sliding ferroelectricity with different translation operations ($\hat{\tau}_z$ is chosen as negative in the work). In Figure. 1(b) and (c), we calculate the out-of-plane ($P_\perp$) and in-plane polarization ($P_\parallel$) in *h*-BN bilayer in configuration space. Consistent with the conclusion in Figure. 1(d), the findings show that the $P_\parallel$ and $P_\perp$ have the same magnitude and cannot be ignored. The opposite $P_\perp$ exists in *M* and *N* points, and only $P_\parallel$ is found in boundary and diagonal lines (*G-G* paths). The maximum $P_\parallel$ occurs at *A*, *B*, and *C* points.

The moiré supercell can be approximated as a combination of bilayer systems with varying translation operations. We can predict the polarization pattern in moiré superlattice by fitting local ferroelectric polarization with machine learning. From this, we hope the polarization can be well defined for each unit cell or even atom. Therefore, MLWFs are a suitable choice to calculate polarization compared with the berry phase method. The WCs of all valence electrons are projected through MLWFs, which can be regarded as the electron coordinate positions in real space, and the atom coordinates can be regarded as the positive charge positions. All positions of the positive and negative charges can be achieved and the polarization can be calculated.

In *h*-BN bilayer, the electronegativity of N atom is greater than B atom and combined with the electron configuration in the pseudopotential of the B and N atoms ($s^2p^1$, $s^2p^3$), it can be assumed that all eight electrons are distributed around the N atom, the MLWCs projection using $sp^3$ orbitals of N atoms is chosen to obtain high-quality WCs. After assigning a charge of 2e⁻ to the WCs, the polarization is the dipole moment of the point charge neutral system composed of the WCs and atomic nucleus. For *h*-BN bilayer, each N atom is assigned a local dipole moment *p*. Specifically, *p* is a weighted contribution from B and N atoms in the same primitive cell, expressed as:

$$\boldsymbol{p}_{i\sim j} = 3\sum_{i\sim j} B_{site} + 5\sum_{i\sim j} N_{site} - 2\sum_{i\sim j} WC_{site} \pm nN \tag{1}$$

where $B_{site}$, $N_{site}$ and $WC_{site}$ are the positions of B, N atoms and WCs, respectively, and *nN* represents an arbitrary polarization quantum. Figure. 1(e) shows how to calculate polarization using WCs, and we can sum and average WCs to get the wannier centroid in N atoms, and *p* of the up and down layer and the global dipole moment also can be defined. From the MLWFs perspective, the sliding ferroelectricity of bilayer system can be viewed as the difference between the dipoles of bilayers. In the $\{E|N\}$ bilayer, the local polarization of the up and down layer ($P_{layer\_up}$, $P_{layer\_dw}$) is opposite in direction but not equal and finally produces a downward net polarization like Figure. 1(e). In the $\{E|G\}$ and $\{I|\tau_o\}$ bilayer, $P_{layer\_up}$ and $P_{layer\_dw}$ are equal in size and opposite in direction, and there is no net polarization, which might be seen as antiferroelectricity.

We begin to construct the data set. Starting with the AA stacked $\sqrt{3} \times \sqrt{3} \times 1$ *h*-BN bilayer, we apply sliding displacements along the lattice vector and extracted 10 sliding positions, resulting in 100 data points. Additionally, different biaxial strains are sampled to generate 500 data points, of which 450 are used for the training set and 50 for the validation set. We then carry out first-principles calculations by Vienna Ab initio Simulation Package (VASP)[41, 42] with the Projector Augmented Wave (PAW) method, and apply the Generalized Gradient Approximation (GGA) and the Perdew-Burke-Ernzerhof (PBE) exchange-correlation functional[43, 44]. The plane-wave energy cutoff is set to 600 eV, and Γ-centered 24×24×1 k points grid is adopted for Brillouin zone sampling. At last, WCs are obtained by Wannier90 software package[45].

The DW model for *h*-BN bilayer is trained using the DP-JAX[38]. The loss function was defined as:

$$Loss = \frac{1}{3N}\sum_i |\Delta\boldsymbol{Dipole}_i|^2 \tag{2}$$

where Δ***Dipole*** represents the difference between the model predictions and the actual dipole values from the training data in each direction. The DP-JAX training was conducted over 100,000 steps, with the learning rate decaying exponentially from $10^{-2}$ to $3.5\times10^{-8}$. Figure. 1(f) shows the prediction accuracy of the DW model compared to the density functional theory (DFT) data (AA stracked bilayer without electrical field). Mean absolute errors (MAE) can be on the order $10^{-5}$ e·Å which is much smaller than $10^{-3}$ e·Å of sliding ferroelectricity of $h$-BN bilayer. In addition, we conducted separate training for the external electric field and the AA' stacked bilayer, and they can achieve the same accuracy.

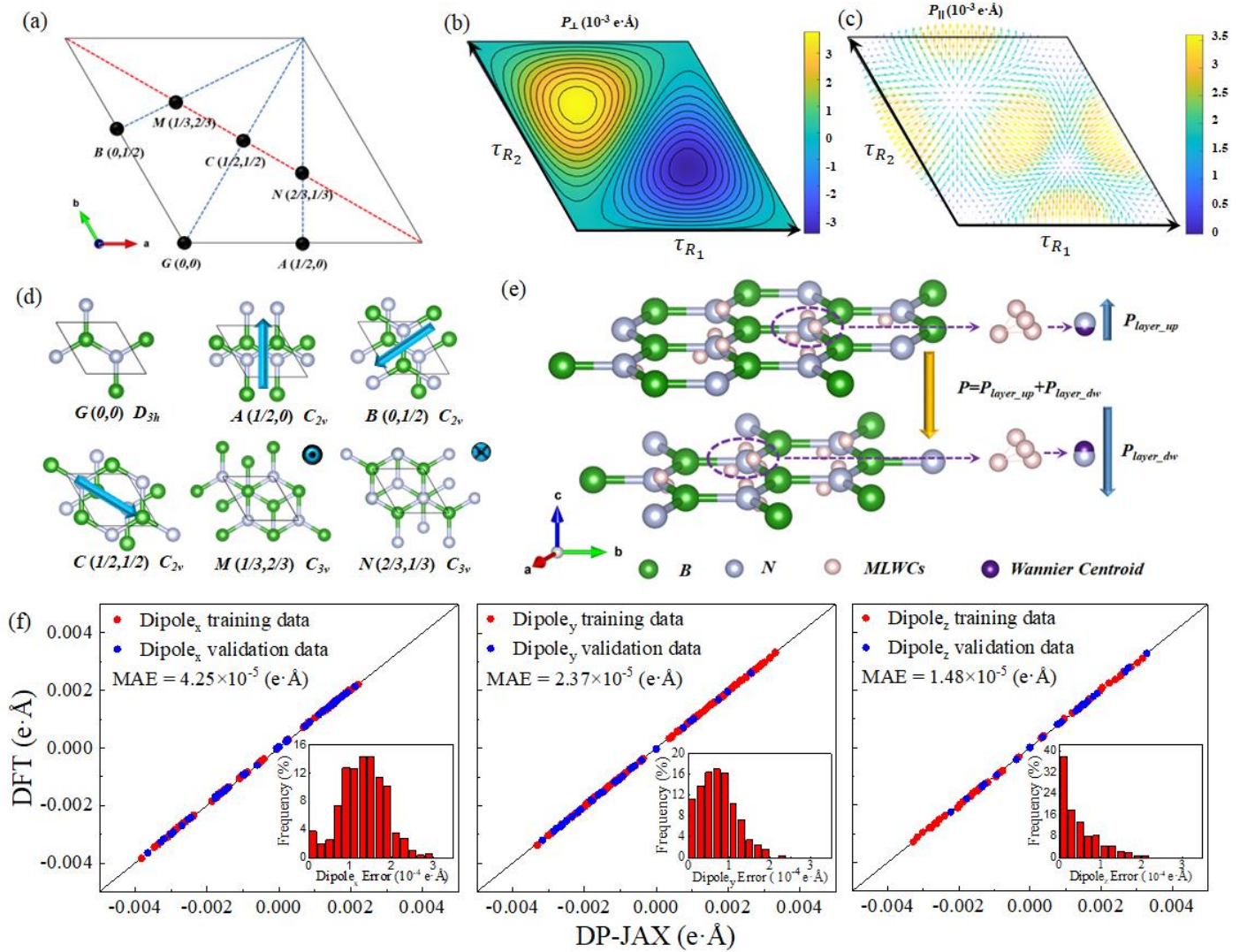

Figure. 1. (a) and (d) are the diagrams of the sliding structure, point group and the direction of sliding ferroelectricity. (b) and (c) are the out-of-plane and in-plane polarization pattern in configuration space $\tau(R_1, R_2)$, where $R_1$ and $R_2$ represent the vector of $h$-BN bilayer primitive. (e) the schematic diagram of calculating polarization by WCs. (f) Comparison of dipole from the DW model and DFT calculations (AA stacked bilayer without electrical field).

To verify the validity of the DW model, we directly calculate the polarization of $\theta = 6.01°$ twisted bilayer by WCs. Figure. 2(a) depicts the crystal structure, while Figure. 2(b) shows the polarization on each N atoms of up and down layers. Overall, the $P_{\text{layer\_up}}$ and $P_{\text{layer\_dw}}$ exhibit a vortex distribution and have nearly opposite helical phases. The difference in the value and phase of $P_{\text{layer\_up}}$ and $P_{\text{layer\_dw}}$ constitute the polarization pattern of the twisted bilayer. From the local perspective, the sum of the polarizations of the two N atoms closest to each other in the up and down layers can be used as the local polarization near the $G$ point. However, the three adjacent atoms determine the local polarization near the $M$ and $N$ points, which is different from the sliding primitive cell. This can be attributed to the lack of translational symmetry of the local position, and the rotation operation is not fully equivalent to the translation operation of the sliding primitive cell. The polarization pattern is displayed in Figure. 2(c), from which it can be seen that the $M$ and $N$ points show vortex with opposite phases and have opposite maximum $P_\perp$. Only $P_\parallel$ on the boundary and diagonal lines of the twisted bilayer, and the maximum $P_\parallel$ is located the $A$, $B$, and $C$ points. In general, although the interlayer rotation $\theta$ is larger, Figure. 2(c) still shows a typical moiré polarization pattern, but the role of surrounding atoms must be considered. The results of model prediction and first-principles calculation are basically consistent (MAE = $5.04 \times 10^{-5}$ e·Å), which proves the validity of the model.

Next, we chose a larger twisted bilayer (1.54°) to discuss the result of model prediction. The in-plane and out-of-plane polarization pattern in Figure. 2(d) and (e) are more accurately mapped to the sliding primitive cell as the $\theta$ decreases. The intricate winding that is topologically nontrivial, such as skyrmions in magnetic systems and ferroelectric domains[27, 46], which are more in line with the results of analytical form and continuous model. In more detail, two triangular regions surrounded by boundary and diagonal lines each have a half-skyrmion (antiskyrmion), that is, merons (antimerons). To prove this, we project the periodic polarization pattern to the sphere and calculate the local topological charge ($q$), whose integral as winding number ($Q$) describes different topological structures. The formula is as follows:

$$Q = \int q \, d\mathbf{r} = \frac{1}{4\pi} \int \overline{\mathbf{P}} \cdot \left( \frac{\partial \overline{\mathbf{P}}}{\partial x} \times \frac{\partial \overline{\mathbf{P}}}{\partial y} \right) d\mathbf{r} \tag{3}$$

where $\overline{\mathbf{P}}$ is normalized polarization and $\mathbf{r}$ is the coordinate in real space. The local topological charge is shown in Figure. 2(f) and it is concentrated near $M$ and $N$ points and is zero along the $G$-$A$/$B$/$C$ path (boundary and diagonal lines). We sum the triangular regions centered at $N$ and $M$ points, $Q_{(M, N)} = \pm 1/2$ and $Q$ of the total twisted bilayer is zeros. Thus, the polarization pattern of the twisted bilayer can be viewed as a combination of Bloch-type merons and antimerons, which corresponds to the polarization of curling around $N$ and $M$ points.

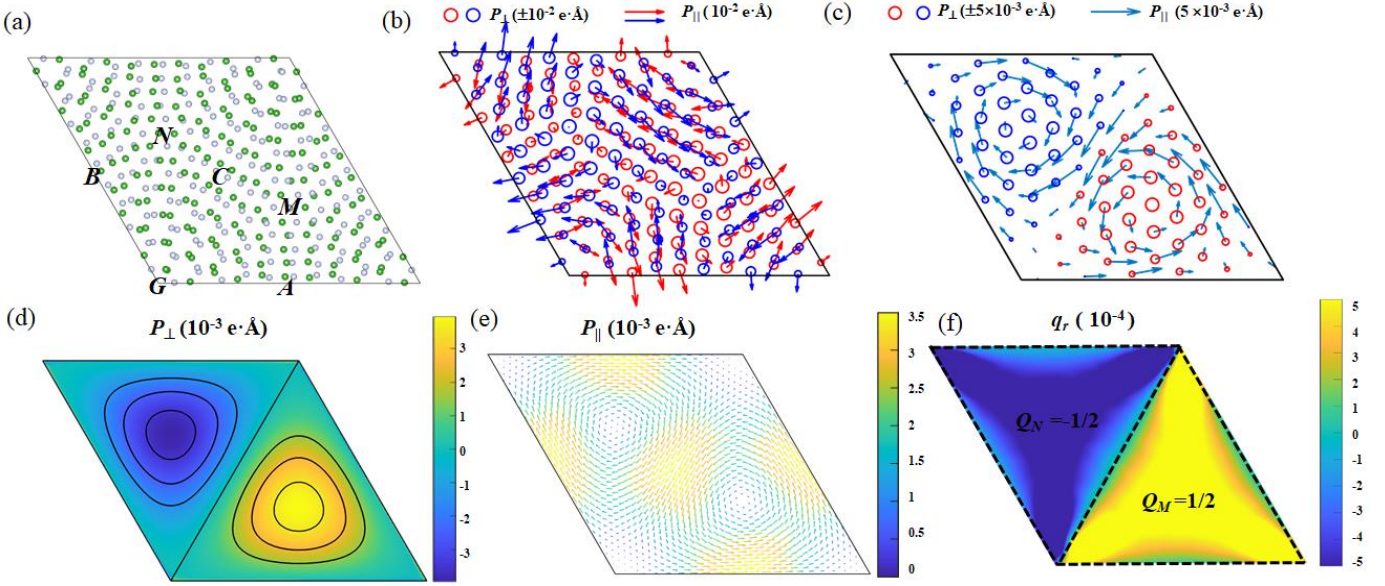

Figure. 2 The crystal structure (a) and polarization distribution of each N atoms (b) of 6.01° twisted bilayer. The red and blue represent the up and down layers, and the $P_\perp$ of upper (lower) layer is always upward (downward). (c) The corresponding polarization pattern. The red and blue circles represent the upward and downward polarization and the diameter represents the magnitude of the $P_\perp$. The out-of-plane (d), in-plane polarization pattern (e), and the local topological charge (f) of 1.54° twisted bilayer. The dashed triangle box indicates the triangular region centered at $M$ ($N$) point.

In bilayer system, the lattice mismatch can also cause long-period moiré patterns, which also can be mapped to the sliding primitive cell. In Figure. 3(a), we display the moiré supercell with 10% lattice mismatch to show the structure clearly. More specifically, the biaxial strain ($\eta$) is applied to the lower layer (compression (tension) strain is + (-)). In Figure. 3(b), the polarization of each N atom is given, and the most important difference between twisted and strained bilayers is the additional unpaired N atoms in the *A-N-C-M-B* path. The polarization pattern of $\eta$ =5% strained bilayer is given in Figure. 3(c). Due to the mapping relationship, its polarization pattern should be the same as that in Figure. 1(b) and (c), but the results differ. The $P_\parallel$ is consistent, but $P_\perp$ is not, which can be seen as out-of-plane uniform polarization applied to Figure. 1(b). The origin of extra polarization is also the lattice mismatch. In the case of the $G$ point, mirror symmetry in the primitive cell makes out-of-plane polarization nonexistent. However, lattice mismatch breaks the local mirror symmetry and produces additional $P_\perp$, which is ignored in effective models or direct mappings. As $\eta > 0$, the additional out-of-plane polarization decreases and the polarization pattern is more consistent with Figure. 1(b). Figure. 3(d) shows the out-of-plane polarization pattern of $\eta = 1\%$ strained bilayer, where the dashed line represents the contour line of $P_\perp = 0$. The dashed line in Figure. 3(d) is closer to the boundary and diagonal lines than Figure. 3(c) indicates the reduction in additional $P_\perp$, consistent with the previous conclusions.

Next, we discuss the topological structure of the polarization pattern in strained bilayer. Consider the ideal case where $\eta > 0$, the out-of-plane polarization pattern of strained bilayer and twisted bilayer is the same, and the in-plane polarization has a 90° phase difference, which does not change $q(r)$ and $Q$. For the strained bilayer, polarization flows into and out the $M$ and $N$ points, which can be viewed as Néel-type meron and antimeron ($Q_{(M, N)} = \pm1/2$). Interestingly, $Q_{(M, N)} = \pm1/2$ is a critical state, where a small change in additional $P_\perp$ causes meron or antimeron to disappear, producing an antiskyrmion or skyrmion. When additional $P_\perp$ is 0, the $P_\perp$ around $N$ and $M$ points is opposite, the helical phase of $P_\parallel$ is opposite, and only $P_\parallel$ exists at the boundary of the triangular region, which just separates the meron and antimeron. When additional $P_\perp$ is present, the $P_\perp$ appears at the boundary, which wraps a complete skyrmion or antiskyrmion. In $\eta = 1\%$ strained bilayer, upward polarization occurs around the $N$ point, and the triangular region corresponding to the $N$ point wraps up the complete antiskyrmion ($Q_N = -1$), while the meron of the $M$ point disappears ($Q_M = 0$), as shown in Figure. 3(e). The Figure. 3(f) shows $q(r)$ in $\eta = -1\%$ strain bilayer. In strained bilayer with tensile strain, the in-plane polarization pattern is the same as compressive strain, the directions of $P_\perp$ and additional $P_\perp$ are switched. This causes the $M$ and $N$ positions to be switched, and the triangular region centered at $M$ point wraps up the complete skyrmion ($Q_M = 1$), while the antimeron of the $N$ point disappears ($Q_N = 0$).

Additionally, there is an interesting issue to discuss in strained bilayer. The upper and lower layers have the same number of primitive cells in twisted bilayer, which can always be mapped back to the sliding primitive bilayer one by one. However, it is not the same in strained bilayer, suggesting that there are locations, which along $A$-$N$-$C$-$M$-$B$ path, cannot be mapped. These results in the strained bilayer always having a net polarization even if $\eta \to 0$, which is always directed from the layer with fewer primitive cells to the layer with more primitive cells, and there may be unique electron accumulation along the $A$-$N$-$C$-$M$-$B$ path.

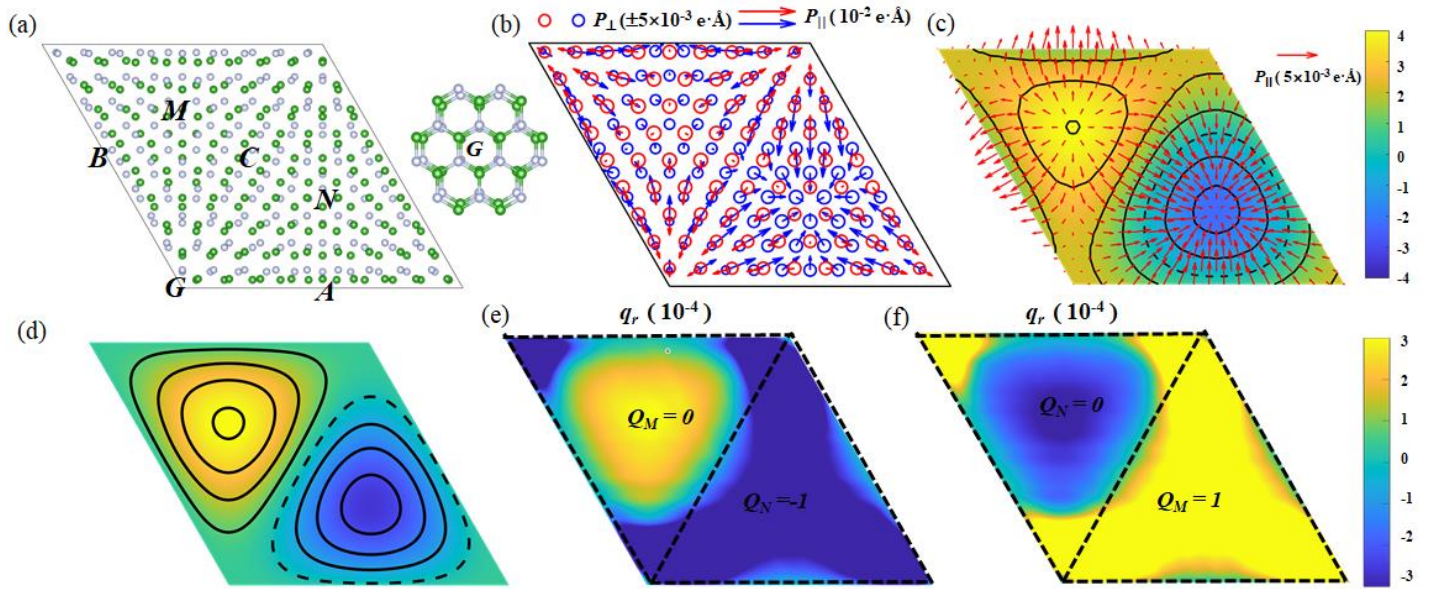

Figure. 3 The crystal structure (a), the polarization distribution of each N atoms (b) of $\eta = 10\%$ strained bilayer. (c) The polarization pattern of $\eta = 5\%$ strained bilayer. (d) The out-of-plane polarization of $\eta = 1\%$ strained bilayer. Dashed lines represent contours with $P_\perp = 0$. The local topological charge of $\eta = 1\%$ (e) and -1% (f) strained bilayer.

We note that additional $P_\perp$ caused by lattice mismatch can change the $q(r)$ and $Q$, suggesting that the external electric field has the opportunity to manipulate the topological structure of the polarization pattern. The Figure. 4(a) and (d) display the $P_\perp$ and $P_\parallel$ along $G$-$N$-$C$-$M$-$G$ path with external electric fields. The response of $P_\perp$ is much greater than $P_\parallel$, which means that the $P_\parallel$ is difficult to adjust by z-direction electric field. The out-of-plane polarization pattern and the local topological charge of 1.085° twisted bilayer with 0.02 V/Å are shown in Figure. 4(b) and (e), respectively. Compared with the zero electrical situation, the whole $P_\perp$ is reduced and the zero contours shrink to $M$ points. The meron of triangular region centered at $M$ point turns into the skyrmion ($Q_M = 1$) and the antimeron of the $N$ point vanishes ($Q_N = 0$), as shown in Figure. 4(e). Continue increasing the electric field until all the $P_\perp$ are flipped in the same direction (0.03 V/Å), and the corresponding polarization pattern and local topological charge are shown in Figure 4(c) and (f). All $P_\perp$ are in the same direction, causing $Q_M$ and $Q_N$ to be 0, and the topological structures disappear.

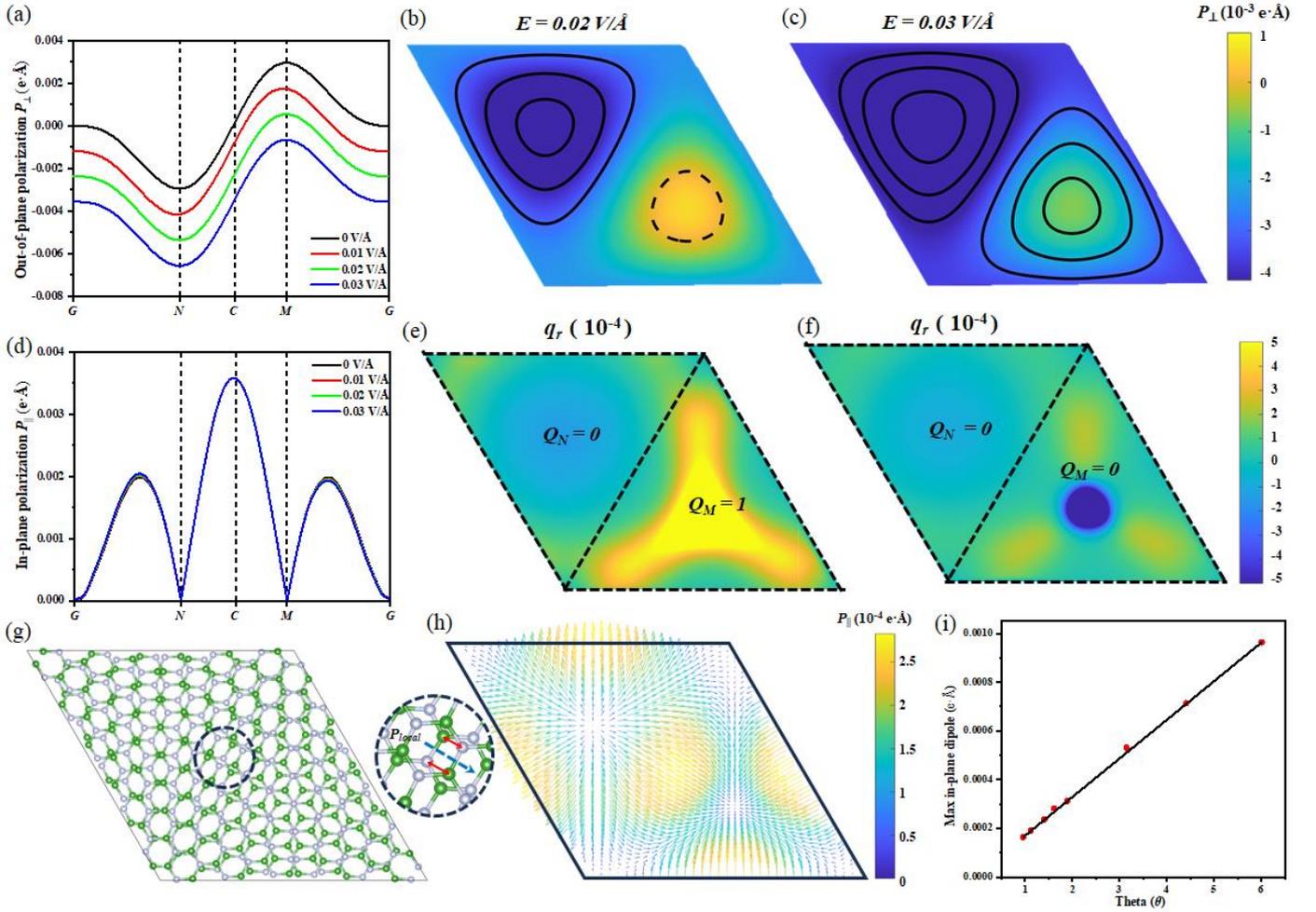

Figure. 4 The $P_\perp$ (a) and $P_\parallel$ (d) along the $G$-$N$-$C$-$M$-$G$ path. The out-of-plane polarization pattern of 1.085° twisted bilayer with 0.02 (b) and 0.03 V/Å (c). The local topological charge of 1.085° twisted bilayer with 0.02 (e) and 0.03 V/Å (f). (g) The AA' stacked 3.15° twisted bilayer crystal structure. (h) The polarization pattern of AA' stacked 1.614° twisted bilayer. (i) The max in-plane dipole vs $\theta$. in AA' stacked twisted bilayer.

Finally, we intend to illustrate the polarization pattern of the antiparallel stacked (AA' stacked) twisted bilayer, which is useful to demonstrate the role neglected by the effective model or direct mapping. Since the sliding operations do not break $I$, the AA' stacked bilayers do not have polarization under any translation operation. However, for finite systems, rotational operation is not fully equivalent to translation operation due to the lack of translation symmetry, which results in polarization, as shown in Figure. 4(g). We give its polarization pattern in Figure. 4(h), its $P_\perp$ is 0, and the in-plane polarization pattern is similar to Figure. 1(c), but an order of magnitude smaller. The origin of this $P_\parallel$ is the absence of local translational symmetry. Therefore, $P_\parallel$ also becomes smaller with the decrease of $\theta$, as shown in Figure. 4(i).

In general, the topological structure of polarization pattern in twisted and strained $h$-BN bilayers is obtained via deep Wannier. The detailed discussion is given to the additional polarization brought on by the absence of local translational symmetry is discussed in detail, which can even be seen in antiparallel stacked twisted bilayer and usually be disregarded in efficient model and direct mapping methods. Based on this, we investigate how the topological structure of polarization pattern is affected by lattice mismatch and external electric field. The polarization pattern consists of meron and antimeron occupying two triangular regions in moiré supercell when there are no external electric fields or lattice mismatch. The combination of meron and antimeron is the critical case, and meron (antimeron) immediately transforms into a complete skyrmion (antiskyrmion) upon applying out-of-plane extra polarization, whether it is from applied electrical field or lattice mismatch. The winding number becomes 0 and the topological structures vanish when all out-of-plane polarizations point in the same direction. Our work provides a more detailed description of the pattern and modulation of dipoles in the moiré systems, which can deepen the understanding of moiré ferroelectricity and promote the development of sliding ferroelectric and topological physics.

**ACKNOWLEDGMENTS**

This work was supported by the National Key Research and Development Program of China (Grants No. 2022YFA1402902 and 2021YFA1200700), the National Natural Science Foundation of China (Grants No. 12134003), China National Postdoctoral Program for Innovative Talents (Grant No. GZC20230809), Shanghai Science and Technology Innovation Action Plan (No. 21JC1402000), ECNU Multifunctional Platform for Innovation.